\documentclass[11pt,letterpaper]{amsart}

\usepackage{amssymb}
\usepackage{url}
\usepackage{graphicx}     
\usepackage{color}

\newtheorem{Thm}{Theorem}
\newtheorem{Lem}[Thm]{Lemma}
\newtheorem{Cla}[Thm]{Claim}
\newenvironment{Proof}{\begin{proof}}{\end{proof}}

\newcommand{\GF}{\ensuremath{\mathrm{GF}}}
\newcommand{\bra}{\ensuremath{\langle}}
\newcommand{\ket}{\ensuremath{\rangle}}
\newcommand{\dtext}[1]{{\em\bf {#1}}}

\begin{document}


\title%
[]%
{Barriers and Local Minima in Energy Landscapes\\
of Stochastic Local Search}
\author%
[]%
{\vspace*{-0.15in}Petteri Kaski}
\address{%
Helsinki Institute for Information Technology HIIT,
Department of Computer Science, University of Helsinki,
P.O.Box 68, FI-00014 University of Helsinki, Finland}
\email{petteri.kaski@cs.helsinki.fi}

\newcommand{\abst}[0]{%
\vspace*{-0.1in}
A local search algorithm operating on an instance
of a Boolean constraint satisfaction problem 
(in particular, $k$-SAT) can be viewed as a stochastic process 
traversing successive adjacent states in an ``energy landscape'' 
defined by the problem instance on the $n$-dimensional Boolean 
hypercube.
We investigate analytically the worst-case topography of such 
landscapes in the context of satisfiable $k$-SAT via a random 
ensemble of satisfiable ``$k$-regular'' linear equations modulo 2.

We show that for each fixed $k=3,4,\ldots$, the typical 
$k$-SAT energy landscape induced by an instance drawn from the 
ensemble has a set of $2^{\Omega(n)}$ local energy minima, 
\emph{each} separated by an unconditional $\Omega(n)$ energy barrier 
from each of the $O(1)$ ground states, that is, solution states 
with zero energy. 
The main technical aspect of the analysis is that a random
$k$-regular 0/1 matrix constitutes a strong boundary expander 
with almost full $\GF(2)$-linear rank, a property which also 
enables us to prove a $2^{\Omega(n)}$ lower bound for the expected 
number of steps required by the focused random walk heuristic 
to solve typical instances drawn from the ensemble.
These results paint a grim picture of the worst-case topography
of $k$-SAT for local search, and constitute apparently the first 
rigorous analysis of the growth of energy barriers 
in a random ensemble of $k$-SAT landscapes
as the number of variables $n$ is increased.
}


\begin{abstract}
\abst
\end{abstract}

\vspace*{-0.4in}
\maketitle
\vspace*{-0.3in}


\section{Introduction}

\subsection{Background and Motivation}
\label{sect:background}
Stochastic local search algorithms \cite{AaLe97,HoSt05} have 
in practice proven to be surprisingly efficient in solving 
instances of difficult constraint satisfaction 
problems (see \cite{ArAu06,SeAO05} for recent examples).
Yet the basic analytical principles underlying the success or failure 
of local search heuristics are far from being understood.

The objective of the present work is to shed new analytical light 
into the combinatorial phenomena that can occur in 
``energy landscapes'' \cite{ReSt02} governing the operation of 
most local search algorithms used in practice. Indeed, the 
difficulty in analyzing even the most elementary heuristics largely 
stems from the fact that the energy landscapes induced by the problem 
instances do not easily yield to combinatorial analysis.

To set the stage, certainly among the most well-understood
settings for constraint satisfaction problems is a system of linear 
equations $Ax\equiv b\pmod 2$ over $n$ variables $x_1,x_2,\ldots,x_n$ 
assuming 0/1 values (that is, ``XORSAT'').
The following example will provide to be illustrative.
\begin{equation}
\label{eq:example-system}
\hspace*{0.4in}
\left[\begin{array}{cccc}
0&1&1&1\\
1&0&1&1\\
1&1&0&1\\
1&1&1&0
\end{array}\right]
\left[\begin{array}{c}
x_1\\
x_2\\
x_3\\
x_4
\end{array}\right]\equiv
\left[\begin{array}{c}
0\\
0\\
0\\
0
\end{array}\right]\pmod 2.
\end{equation}

A local search algorithm can now be viewed as a stochastic 
process that traverses a sequence of adjacent states in the 
\dtext{energy landscape} associated with the problem instance.
For a linear system $Ax\equiv b\pmod 2$, the \dtext{states} of the 
landscape consist of the $2^n$ possible
assignments $s=(s_1,s_2,\ldots,s_n)$ of 0/1 values to the
variables $x_1,x_2,\ldots,x_n$.
Any two states are \dtext{adjacent} if they differ in the value of 
exactly one variable; the \dtext{distance} between two states is
the number of variables having different values in the two states.
Associated with each state $s$ is an \dtext{energy} $E(s)$ equal 
to the number of equations violated by the assignment $x=s$. 
For example, with lines indicating adjacency and energy indicated 
by subscripts, the landscape associated with (\ref{eq:example-system}) 
is depicted below.
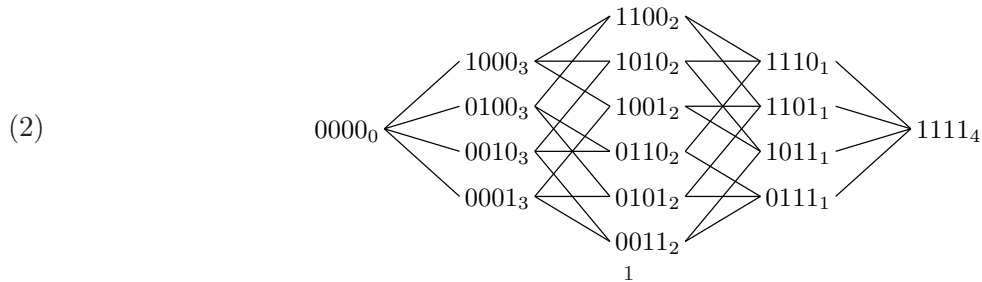
\begin{equation}
\label{eq:example-landscape}
\begin{minipage}{6in}
\begin{center}
\input{example-landscape.input}
\end{center}
\end{minipage}
\end{equation}

The ``simple'' setting of linear equations 
is motivated because it provides direct insight into 
landscape phenomena in less tractable settings, in particular, 
in the context of the 
\dtext{$k$-satisfiability problem ($k$-SAT)} \cite{DuGP97}.
Indeed, a linear equation with $k$ variables is logically equivalent 
to a conjunction of $2^{k-1}$ SAT clauses of length $k$ that exclude the 
0/1 assignments violating the equation. Furthermore, assuming that energy
in SAT is defined as the number of violated clauses, the landscape
of the SAT encoding of $Ax\equiv b\pmod 2$
is \emph{identical} to the linear landscape.
Thus, any landscape phenomenon that occurs in the context of linear
equations also occurs in SAT.

In the present work we seek to understand what an energy 
landscape ``can look like'' to local search 
heuristics, in the worst case.
The two standard heuristics that occur in most local 
search algorithms are:
(a) \dtext{energy bias}---the algorithm prefers (in probability) moving 
into adjacent states with lower energy over those with higher energy; and 
(b) \dtext{focusing}---the algorithm prefers moving into adjacent 
states such that the move affects the constraints that are violated 
in the current state.

Exerting an energy bias does not always guide a search 
towards a solution, as can be immediately seen from
(\ref{eq:example-landscape}).
To study the worst-case extent of this phenomenon, 
we consider two standard combinatorial measures of ``ruggedness'' in
a landscape:
(a) the \dtext{local minimum} states, 
that is, the states with positive energy whose adjacent states all 
have strictly higher energy, and
(b) the global \dtext{energy barrier} separating 
a state $s$ from a state $t$, that is, the minimum increase in 
energy over $E(s)$ required by \emph{any} walk from $s$ to $t$ 
consisting of successive adjacent states.
Of special interest are the barriers separating local minima from 
\dtext{ground states}, that is, the zero-energy solution states.
For example, in (\ref{eq:example-landscape}) the local minimum
states are $1110$, $1101$, $1011$, and $0111$, each 
separated by a barrier of $3-1=2$ from the unique ground state $0000$.

From the perspective of the focusing heuristic,
a benchmark algorithm is the \dtext{focused random walk} \cite{Papa91} 
(in each step, select uniformly at random one violated constraint,
and flip the value of one variable selected uniformly at random
among the variables occurring in the constraint).
Also focusing can perform poorly, as can be seen by
considering the transition probabilities 
in (\ref{eq:example-system}) and (\ref{eq:example-landscape})
for the focused random walk.

The subsequent analysis paints a grim picture of the worst-case 
topography that heuristics face already in the ``simple'' case 
of $k$-regular linear equations, and hence, in the case of $k$-SAT. 
The present results constitute apparently the first rigorous 
topographical analysis of the energy landscapes induced by a 
nontrivial random ensemble. 
(See \S\ref{sect:related} for a discussion of related work.)

\subsection{Statement of Results}
Throughout this work we assume that $k=3,4,\ldots$ is fixed.
In particular, any asymptotic notation 
$O(\cdot)$, $\Omega(\cdot)$, $o(\cdot)$ always refers to the parameter 
$n$ growing without bound and $k$ remaining fixed.
Furthermore, the constants hidden by the asymptotic notation in 
general depend on the fixed parameters, such as $k$ and $\epsilon$ in 
Theorem \ref{thm:main}.

An $n\times n$ matrix with 0/1 entries 
is $k$-\dtext{regular} if every row and every column has exactly 
$k$ nonzero entries. For a given $n$, a \dtext{random $k$-regular matrix} 
refers to a $k$-regular $n\times n$ matrix selected uniformly at random 
from the set of all such matrices. Similarly, 
a \dtext{random $k$-regular landscape} refers to the energy landscape
associated with a system $Ax\equiv 0\pmod 2$, where $A$ is
a random $k$-regular matrix.

\begin{Thm}[Energy barriers and local minima]
\label{thm:main}
For each fixed $k=3,4,\ldots$ and $\epsilon>0$ it holds that
a random $k$-regular landscape has with probability
at least\/ $1-\epsilon$ the following three properties:
\begin{itemize}
\item[(i)] 
the number of ground states is $O(1)$;
\item[(ii)]
any two distinct ground states have distance $\Omega(n)$
and are separated by an $\Omega(n)$ energy barrier from each other;
\item[(iii)] 
there exists a set of\/ $2^{\Omega(n)}$ local minima such that
each local minimum is separated by an $\Omega(n)$ energy 
barrier from every ground state.
\end{itemize}
\end{Thm}
Thus, an energy landscape can be very uneven indeed.
Furthermore, Theorem \ref{thm:main} leaves no possibility 
for ``trivial'' barriers caused by large local fluctuations of energy. 
Indeed, because each variable occurs in $k=O(1)$ equations,
it follows that moving from one state into an adjacent state changes 
the energy by at most $k$ units, implying that the extensive 
energy barriers are a global phenomenon apparently not easily 
circumvented with local heuristics. 
Due to the connection with $k$-SAT, identical lower bounds hold 
for $k$-SAT landscapes in the worst case. Interestingly, this worst-case 
phenomenon occurs at a ratio $\alpha=2^{k-1}$ of clauses to variables, 
which is well below the SAT/UNSAT threshold \cite{AcPe04,Frie99,MeMZ05a} 
for the ``random $k$-SAT'' \cite{ChKT91,ChSz88,MiSL92} ensemble.

Also the focused random walk can be shown to fail systematically 
for random $k$-regular systems.

\begin{Thm}[Lower bound for focused random walk]
\label{thm:main2}
For each fixed $k=6,7,\ldots$ and $\epsilon>0$ it holds that
the system $Ax\equiv 0\pmod 2$ defined by a random 
$k$-regular matrix $A$ has with probability
at least\/ $1-\epsilon$ the property 
that the focused random walk
requires\/ $2^{\Omega(n)}$ expected steps to arrive at a ground state
when started from an initial state selected uniformly at random.
\end{Thm}

The main technical hurdle in establishing Theorems \ref{thm:main} 
and \ref{thm:main2} is the following result, which we expect to be 
of independent interest (see \S\ref{sect:related}) in particular due 
to its role in establishing the existence of strong $k$-regular boundary 
expanders with almost full linear rank.
\begin{Thm}
\label{thm:kernel}
The expected size of the kernel of a random $k$-regular matrix 
over\/ $\GF(2)$ is $O(1)$.
\end{Thm}

A matrix $A$ is a $(k,\omega,\eta)$-\dtext{boundary expander} if
(a) the number of nonzero entries in every column is at most $k$,
and 
(b) for all $w=1,2,\ldots,\lfloor\omega\rfloor$, every submatrix 
    consisting of $w$ columns of $A$ has at least $\lceil\eta w\rceil$ rows 
    containing exactly one nonzero value.
The following theorem is well known (cf.~\cite[Theorem~4.16(2)]{HoLW06}).

\begin{Thm}
\label{thm:expansion}
For each fixed $k=3,4,\ldots$ and $\delta>0$ 
there exists a $\beta>0$ such that a random $k$-regular matrix
is a\/ $(k,\beta n,k-2-\delta)$-boundary expander with 
probability\/ $1-o(1)$. 
\end{Thm}

\vspace*{-\smallskipamount}

\textbf{[[} N.B.~A proof of 
Theorem \ref{thm:expansion} is provided in 
Appendix \ref{sect:expansion-proof}. 
\textbf{]]}

\medskip

Applying Markov's inequality to Theorem \ref{thm:kernel} 
and combining with Theorem \ref{thm:expansion}, it follows 
that for each fixed $k=3,4,\ldots$, $\delta>0$, and 
$\epsilon>0$ there exist constants $d>0$ and $\beta>0$ such that
with probability at least $1-\epsilon$ a random $k$-regular matrix both 
(a) has a kernel of size at most $2^d$
and 
(b) is a $(k,\beta n,k-2-\delta)$-boundary expander.
This provides the technical foundation for 
Theorems \ref{thm:main} and \ref{thm:main2}.

\subsection{Connections and Related Work}
\label{sect:related}
Random ensembles of constraint satisfaction problems such as 
``\dtext{random $k$-XORSAT}'' \cite{CrDa99,RiWZ01,Scha78} and 
``\dtext{random $k$-SAT}'' \cite{ChKT91,ChSz88,MiSL92}
have received extensive attention both from the computer science 
and the statistical physics communities 
\cite{AcNP05,DMSZ01,GoSe05,HoHW96,KiSe94,MePZ02,MZKS99}. 
In particular, the random $k$-XORSAT ensemble is by
now well-understood as regards rigorous analysis of the transition 
phenomena as the ratio $\alpha$ of the number of equations to variables 
is increased \cite{CoDM03,CrDa99,DuMa02,MeRZ03}, and a similar rigorous 
foundation is emerging for random $k$-SAT 
\cite{AcPe04,AcRi06,Frie99,MeMZ05b},
where the corresponding control parameter $\alpha$
is the ratio of the number of clauses to variables. 
The present work differs from these studies by
(a) considering an essentially different random ensemble, and
(b) focusing on the topography of the complete energy landscape, 
whereas most of the recent effort, 
e.g.~\cite{AcRi06,MeMZ05b,MePR05,MoMe06,MoMZ05}, in studies 
of random $k$-XORSAT and random $k$-SAT has gone to investigating
``only'' the distance distribution between the ground states akin to 
Theorem \ref{thm:main}(ii). (An exception is \cite{MoSe06a}, where 
it is shown that in the limit $n\rightarrow\infty$ the energy barriers 
in random $k$-XORSAT between nearby ground states are bounded from 
below by $-C\log(\alpha_\mathrm{d}-\alpha)$ for some constant $C>0$ 
as the control parameter $\alpha$ approaches the dynamical transition 
point $\alpha_\mathrm{d}$ \cite{MeRZ03}.)
The growth of energy barriers and local minima 
as a function of the system size $n$ has apparently 
not been rigorously investigated in random ensembles 
until the present work.

The structure of energy landscapes associated with local search
algorithms and spin-glass models of statistical physics 
\cite{BoBo07,MePV87} have been the focus of many empirical and 
quasi-rigorous statistical-physics studies, 
e.g.~\cite{BaHW03,BKMT06,DaSi03,FHSW02,FrCS97,SeMo03}, 
however, rigorous results are more scarce. In this connection at
least one result exists, namely in \cite{NeMo99} it is shown
that a deterministic 3-regular matrix family based on a triangular 
lattice has an associated landscape with local minima separated by an
$\Omega(\log(n))$ barrier from the ground state;
a benchmark study of SAT-solvers using $3$-SAT instances derived
from this family is carried out in \cite{JiMS05}.
A general survey of combinatorial landscapes in various contexts 
is \cite{ReSt02}.

From the perspective of computer science and statistical physics,
the ``random satisfiable $k$-regular XORSAT'' 
(``ferromagnetic $k$-spin model 
with Ising spins and fixed connectivity $k$'') 
ensemble studied in the present work 
has apparently been the focus of only relatively few studies, 
despite the fact that the study of random $k$-regular matrices 
(equivalently, random $k$-regular bipartite graphs with a fixed 
bipartition) has a long history in mathematics \cite{Boll01,Worm99}.
To the best of our knowledge, from a computational / statistical physics 
perspective the few works addressing the present ensemble are
\cite{MoSe06b}, where an analysis of the correlation times of the 
Glauber dynamics on a corresponding spin-glass model is carried out, 
and \cite{HJKN06}, where clausal encodings for the $k=3$ case 
are used to empirically benchmark SAT-solvers; further experiments
for the $k>3$ case are reported in \cite{Jarv06}.
Statistical physics studies on analogous fixed-connectivity models
include \cite{FLRZ01,FMRW01,MoRi04,RiKi92}.

From a mathematical perspective it is immediate that the analysis of 
$k$-regular matrices over $\GF(2)$ is closely related to the study 
of \dtext{low-density parity-check codes (LDPC codes)} 
\cite{Gall63,RiUr06} in coding theory. 
In coding-theoretic language, Theorem \ref{thm:kernel} 
states that the expected total number of codewords 
in a linear code defined by a parity-check matrix drawn from 
the $k$-regular matrix ensemble is $O(1)$ 
(indicating that such codes have very limited applicability
from a coding-theoretic perspective). 
From a methodological perspective, however, the tools used to 
analyze the average weight distribution of the codewords in 
standard LDPC code ensembles are analogous to the tools used 
to prove Theorem \ref{thm:kernel}
(cf.~\cite{BaBu05,BuMi04,DiRU06,LiSh02,OrVZ05,Rath06}),
the main difference being that we want to bound the 
expected \emph{total} number of codewords rather than the
number of codewords with a specific relative weight, 
necessitating uniform upper bounds that enable summation over 
all the weights $w=0,1,\ldots,n$.

Theorems \ref{thm:main} and \ref{thm:main2} are apparently the 
first results where expansion is employed in lower bound results 
aimed at understanding local search, despite the fact that expansion 
is a basic tool in numerous lower bound constructions in, e.g., 
proof complexity \cite{ABRW04,BeWi01,ChSz88,Urqu87}, 
where many constructions are based on clausal encodings of linear 
equations. In particular, the probabilistic full-rank boundary 
expander constructions in \cite{ABBI05,AlHI05} apparently
provide an analogue of Theorem \ref{thm:main} in the special case 
$k=3$; however, this is not immediate due to lack of regularity.
In the converse direction, the present Theorem \ref{thm:kernel}
and Theorem \ref{thm:expansion} imply
(by stripping dependent rows and columns)
the existence of full-rank boundary expanders for every $k\geq 3$,
thereby providing partial progress to the open lower bound
questions in \cite[\S 5]{ABBI05}.
An interesting technical contrast to the present lower bound
results is that the \emph{upper bound} for the focused
random walk in \cite{AlBS03} also relies on typical expansion 
properties of random $3$-SAT instances. 
A recent survey of expansion and its applications is 
\cite{HoLW06}.

A large number of stochastic local search algorithms 
for the $k$-SAT problem are based on variations and combinations 
of the energy bias and focusing heuristics. 
Arguably the two central algorithm families in this respect are
(a) algorithms in the ``WalkSAT family'' \cite{McSK97,SeKC96}
(e.g. \cite{Hoos02,Papa91,Scho02,SeLM92}),
and 
(b) algorithms based on variations of the Metropolis dynamics 
\cite{MeRR53} (e.g.~\cite{ArAu06,Cern85,KiGV83,SeAO05}).
(The recent survey propagation algorithm
\cite{BrMZ05,MePZ02,MoMe02} for random $k$-SAT also employs local 
search, but only as a postprocessing step after a ``global'' form of 
belief propagation \cite{AuGK05,MaMW05}.)

Only relatively few rigorous upper and lower bound results are 
known for the running time of local search algorithms for 
$k$-SAT. For the focused random walk with restarts, 
it is known \cite{Scho02} that a satisfying assignment in 
any satisfiable instance of $k$-SAT is found in $O(n(2-2/k)^n)$ 
expected steps, $k\geq 3$. In \cite{AlBS03} it is shown that
the focused random walk finds a satisfying assignment in 
$O(n)$ steps with high probability for a typical instance
drawn from the random 3-SAT ensemble for $\alpha\leq 1.63$. 
An exponential upper bound improving upon the trivial $O(2^n)$ 
is derived in \cite{Hirs00} for a ``cautious'' randomized greedy 
approach. In terms of lower bounds, families of crafted instances 
whose solution requires an expected exponential number of steps
of the focused random walk 
are known; see \cite{AlBS03} and \cite[\S11.5.6]{Papa94}.
Explicit families of instances forcing exponential expected running 
times for certain randomized greedy heuristics are constructed 
in \cite{Hirs00}. Quasi-rigorous statistical physics studies 
considering local search heuristics include \cite{BaHW03,SeMo03}.

From the perspective of local search algorithms for $k$-SAT,
the present Theorem \ref{thm:main2} apparently provides the first 
example of a nontrivial random ensemble with exponential lower bounds 
on the expected running time for the focused random walk. 
Furthermore, the energy barriers and local minima demonstrated in 
Theorem \ref{thm:main}(iii) constitute a step towards 
rigorous lower bounds for more complex heuristics relying on a 
combination of energy bias and focusing. 
In this regard the subsequent proof of Theorem \ref{thm:main2} 
actually provides a meager first step---for large enough 
$k$ it is immediate from (\ref{eq:drift}) that a comparably ``small'' 
energy bias is insufficient to overcome the systematic drift away 
from ground states caused by focusing and expansion.

As regards energy bias heuristics alone, the convergence 
properties of nonfocused variants of the Metropolis dynamics 
(simulated annealing \cite{Cern85,KiGV83} in particular)
have been extensively analyzed; see 
\cite{AaKo89,Cato99,DiSe05,Haje88,HaCK06,SaSF02}
and the references therein. However, these analyses typically
adopt a worst-case setting necessitating that a ground state 
is found with significant probability from every possible initial state.
To arrive at a rigorous analysis of the typical behavior from a random 
initial state akin to Theorem \ref{thm:main2}, a study of the landscape 
structure beyond the properties in Theorem \ref{thm:main} is apparently 
required. In particular, the structure of the attraction basins 
(see \cite{FHSW02}) of the local minima in Theorem \ref{thm:main}(iii) 
in relation to the attraction basins of the ground states need to be 
better understood.

\subsection{Organization}
The remainder of this work is organized as follows.
The conventions and mathematical preliminaries 
are reviewed in \S\ref{sect:preliminaries}.
Theorem \ref{thm:kernel} is proved in \S\ref{sect:kernel}.
Theorems \ref{thm:main} and \ref{thm:main2} are proved
in \S\ref{sect:topo}. 

\subsection{Acknowledgments}
The author would like to thank Mikko Alava, Pekka Orponen, and Sakari Seitz
for useful discussions, and Jukka Kohonen for insight with the
proof of Lemma \ref{lem:tau}.
This research was supported in part by the Academy of Finland, Grant 117499.

\section{Preliminaries}
\label{sect:preliminaries}

\subsection{Conventions}
A \dtext{vector} always refers to 
an $n$-dimensional column vector with elements in the 
finite field $\GF(2)=\{0,1\}$. All arithmetic on vectors is over $\GF(2)$.
For $j=1,2,\ldots,n$, denote by $e_j$ the \dtext{standard basis} 
vector with the $j$th element equal to $1$ and all other elements equal 
to $0$. A \dtext{state} is a synonym for vector when landscapes are discussed.
The \dtext{weight} $W(u)$ of a vector $u$ is the number of nonzero elements.
In accordance with the definitions in \S\ref{sect:background}, 
the \dtext{energy} of a state $s$ with respect to the system
$Ax\equiv 0\pmod 2$ is defined by $E(s)=W(As)$. 
The \dtext{distance} between states $s$ and $t$ is $D(s,t)=W(s+t)$.
A state $s$ is a \dtext{local minimum} if 
$E(s)>0$ and $E(s+e_j)>E(s)$ holds for all $j=1,2,\ldots,n$.

\subsection{Asymptotics}
All logarithms are to the natural base 
$\exp(1)=\sum_{k=0}^{\infty}1/k!$.
We recall a variant \cite{Robb55} of Stirling's formula,
valid for all positive integers $n$,
\begin{equation}
\label{eq:stirling}
\frac{1}{\sqrt{2\pi}}\exp\biggl(n\log(n)-n+\frac{\log(n)}{2}+\frac{1}{12n+1}\biggr)
\leq n! \leq
\frac{1}{\sqrt{2\pi}}\exp\biggl(n\log(n)-n+\frac{\log(n)}{2}+\frac{1}{12n}\biggr).
\end{equation}
For $0<\lambda<1$, define the entropy function 
$H(\lambda)=-\lambda\log(\lambda)-(1-\lambda)\log(1-\lambda)$.
From (\ref{eq:stirling}) we have the following upper bounds
for the binomial coefficients, valid
for all integers $n,k\geq 3$ and $w=1,2,\ldots,n-1$:
\begin{equation}
\label{eq:binom-bounds}
\binom{n}{w}\leq\exp\biggl(nH\biggl(\frac{w}{n}\biggr)\biggr),
\qquad
\binom{n}{w}\binom{kn}{kw}^{-1}\leq
\sqrt{k}\exp\biggl(-(k-1)nH\biggl(\frac{w}{n}\biggr)+\frac{1}{6kw}\biggr).
\end{equation}

In what follows we require asymptotic approximations for coefficients of 
large powers of certain polynomials. For a polynomial $P(z)$, denote by 
$[z^N]\bigl\{P(z)\bigr\}$ the coefficient of the term $z^N$ in $P(z)$.
For example, $[z^2]\bigl\{1+6z^2+z^4\bigr\}=6$ and 
$[z]\bigl\{1+3z^2\bigr\}=0$.
The following theorem is a well-known ``local limit analogue''
\cite{Bend73,GnKo54} of the central limit theorem in probability theory; 
see \cite[Chap.~IX]{FlSe06}.

\begin{Thm}[Local limit law for coefficients of a polynomial power]
\label{thm:local-limit}
Let $P(z)$ be a polynomial of degree $d\geq 1$ with 
a positive constant term and positive coefficients
such that the greatest common divisor of the degrees of the 
nonzero terms of $P(z)$ is\/ $1$, let 
\[
\mu=\frac{P'(1)}{P(1)},\quad
\sigma^2=\frac{P''(1)}{P(1)}+\mu-\mu^2,\quad
\sigma>0,
\]
and let\/ $0<\delta<2/3$. 
Then, for all large enough $n$, it holds uniformly 
for all integers of the form $N=\mu n+\nu$ with $|\nu|\leq n^{\delta}$
that
\begin{equation}
\label{eq:gaussianpower}
[z^N]\bigl\{P(z)^n\bigr\}=
\frac{1}{\sqrt{2\pi n}\sigma}P(1)^n
\exp\biggl(-\frac{\nu^2}{2\sigma^2 n}\biggr)\bigl(1+o(1)\bigr).
\end{equation}
\end{Thm}

\textbf{[[} N.B.~A proof of 
Theorem \ref{thm:local-limit} is provided in 
Appendix \ref{sect:gaussian-proof}. 
\textbf{]]}

\subsection{The Configuration Model for $k$-Regular Matrices}
\label{sect:configuration}
For integers $n\geq k\geq 3$,
let $X$ and $Y$ be two $kn$-element sets of \dtext{points}, 
both of which are partitioned into $n$ \dtext{cells} of $k$ points each.
A $(k,n)$-\dtext{configuration} is a bijection $\gamma:X\rightarrow Y$.
Denote by $\mathcal{I}$ the set of cells in $X$ and by $\mathcal{J}$ 
the set of cells in $Y$. Associated with a $(k,n)$-configuration 
$\gamma$ there is a $n\times n$ integer matrix $A=(a_{I\!J})$ defined 
for all $I\in \mathcal{I}$ and $J\in \mathcal{J}$ by
$a_{I\!J}=|\{i\in I: \gamma(i)\in J\}|$.
We clearly have $\sum_{J} a_{I\!J}=k$ for all $I\in \mathcal{I}$ and
$\sum_{I} a_{I\!J}=k$ for all $J\in\mathcal{J}$. 
A configuration is \dtext{simple} if $A$ is a 0/1 matrix.

The following theorem is 
due to 
B\'ek\'essy, B\'ek\'essy, and Koml\'os \cite{BeBK72}
and O'Neil \cite{Onei69};
early related results are due to 
Erd\H os and Kaplansky \cite{ErKa46} and Read \cite{Read59}.

\begin{Thm}
\label{thm:simple}
A random $(k,n)$-configuration is simple with probability 
$\mathrm{exp}(-(k-1)^2/2)+o(1)$.
\end{Thm}
It is well known that any given $k$-regular matrix is obtained 
from exactly $(k!)^{2n}$ simple $(k,n)$-configurations, 
enabling one to access the uniform distribution on 
the set of all $k$-regular $n\times n$ matrices via the uniform 
distribution on the set of all simple $(k,n)$-configurations.

Also considerable extensions of Theorem \ref{thm:simple} are known, see 
\cite{Boll01,GoCr77,GrMW06,Worm99}.

\section{Expected Size of The Kernel}
\label{sect:kernel}

We proceed with the proof of Theorem \ref{thm:kernel}.

\begin{Proof}
By linearity of expectation, we can express the expected size of the
kernel as a sum of expectations of 0/1 indicator variables, one
indicator for each of the $2^n$ vectors.
The expectation of each indicator is equal to the 
probability of the corresponding vector occurring in the kernel.
By symmetry, for each weight $w=0,1,\ldots,n$, all the
$\binom{n}{w}$ vectors of weight $w$ have equal probability 
of occurring in the kernel. Denote by $P_k(n,w)$ the probability 
that a given vector $x$ of weight $w$ occurs in the kernel. 

We proceed to derive an upper bound for $P_k(n,w)$ using the 
configuration model. 
We have that $x$ occurs in the kernel of $A$ if and only if
the columns of $A$ corresponding to the $w$ nonzero coordinates of
$x$ form a submatrix with an even number of nonzero entries in every 
row. Let $e_i=0,2,\ldots,2\lfloor k/2\rfloor$ be the number of nonzero
entries in row $i$ of this submatrix. Because $A$ is $k$-regular,
$\sum_{i=1}^n e_i=kw$. 
The number of simple $(k,n)$-configurations that induce an $A$ meeting
a given nonnegative even composition $e_1+e_2+\ldots+e_n=kw$
is at most
$(kw)!\cdot(k(n-w))!\cdot\prod_{i=1}^n\binom{k}{e_i}$.
To obtain an upper bound for the total number of simple $(k,n)$-configurations 
that induce an $A$ with $x$ in the kernel, let
\begin{equation}
\label{eq:ek}
E_k(z)=
\sum_{j=0}^{\lfloor k/2\rfloor}\binom{k}{2j}z^{2j}=\frac{(1+z)^k+(1-z)^k}{2},
\qquad
B_k(n,w)=[z^{kw}]\bigl\{E_k(z)^n\bigr\}. 
\end{equation}
Now observe that the total number of simple $(k,n)$-configurations that 
induce an $A$ with $x$ in the kernel is at most
$(kw)!\cdot(k(n-w))!\cdot B_k(n,w)$,
where $B_k(n,w)$ in effect sums the product $\prod_{i=1}^n\binom{k}{e_i}$
over all the eligible compositions $e_1+e_2+\ldots+e_n=kw$.
By Theorem \ref{thm:simple}, for all large enough $n$ there are at
least $\rho\cdot(kn)!$ simple $(k,n)$-configurations, where $\rho$ is any
positive constant less than $\exp(-(k-1)^2/2)$. 
We thus have the upper bound
$P_k(n,w)\leq \rho^{-1}\binom{kn}{kw}^{-1}B_k(n,w)$.

Taking the sum of $P_k(n,w)$ over all vectors of weight $w$ and all 
weights $w=0,1,\ldots,n$, we have that the expected size of the kernel 
of a random $k$-regular matrix of size $n\times n$ is at 
most $\rho^{-1}S_k(n)$, where
\begin{equation}
\label{eq:sk}
S_k(n)=\sum_{w=0}^n \binom{n}{w}\binom{kn}{kw}^{-1}B_k(n,w).
\end{equation}
The rest of this section provides an asymptotic analysis establishing
that $S_k(n)=O(1)$.
\end{Proof}

\begin{Thm}
\label{thm:sk}
$S_k(n)=2+o(1)$ if $k$ is odd and 
$S_k(n)=4+o(1)$ if $k$ is even.
\end{Thm}
\begin{Proof}
Partition the sum (\ref{eq:sk}) into the following intervals:
\begin{alignat}{2}
\notag
0\leq w&<n/(2k),                  &\qquad &\text{(left extreme deviation)}\\
\notag
n/(2k)\leq w&<(n-n^{3/5})/2,      &\qquad &\text{(left large deviation)}\\
\label{eq:regions}
(n-n^{3/5})/2\leq w&\leq(n+n^{3/5})/2, &\qquad &\text{(central region)}\\
\notag
(n+n^{3/5})/2<w&\leq n(1-1/(2k)), &\qquad &\text{(right large deviation)}\\
\notag
n(1-1/(2k))<w&\leq n.             &\qquad &\text{(right extreme deviation)}
\end{alignat}
Observe that $B_k(n,w)=0$ if $kw$ is odd. Furthermore,
if $k$ is even, we have $B_k(n,w)=B_k(n,n-w)$ by symmetry of the
binomial coefficients, implying that left and right regions are identical 
if $k$ is even. 
If $k$ is odd, then $E_k$ has degree $k-1$, implying 
that $B_k(n,w)=0$ for all $w>(k-1)n/k$ 
and that the sum is zero in the right extreme region.

\begin{Cla}
The sum in the central region is\/ $1+o(1)$ if $k$ is odd 
and\/ $2+o(1)$ if $k$ is even.
\end{Cla}
\begin{Proof}
Using Theorem \ref{thm:local-limit}, we first derive 
Gaussian approximations to the terms $\binom{n}{w}$, 
$\binom{kn}{kw}$, and $B_k(n,w)$ in the central region. 
To this end, let $\delta=3/5$.
From the binomial theorem it follows that 
$\binom{an}{aw}=[z^{aw}]\bigl\{(1+z)^{an}\bigr\}$ 
for nonnegative integers $a,n,w$. 
Setting $P(z)=(1+z)^a$, we have $\mu=a/2$ and $\sigma=\sqrt{a}/2$ 
in Theorem \ref{thm:local-limit}. We obtain that
\begin{equation}
\label{eq:binomial-gaussian}
\binom{an}{aw}=
\frac{1}{\sqrt{2\pi na}}2^{an+1}\exp\biggl(-\frac{a(2w-n)^2}{2n}\biggr)(1+o(1))
\end{equation}
uniformly for all integers $w$ in the central region 
$(n-n^{3/5})/2\leq w\leq (n+n^{3/5})/2$.
To approximate $B_k(n,w)$, let $P(z)=E_k(\sqrt{z})$ 
and observe that $P(z)$ is a polynomial meeting the requirements of Theorem 
\ref{thm:local-limit} with $\mu=k/4$ and $\sigma=\sqrt{k}/4$. 
We obtain
\begin{equation}
\label{eq:bknw-gaussian}
B_k(n,w)=
\begin{cases}
\frac{2\sqrt{2}}{\sqrt{\pi nk}}2^{(k-1)n}\exp\bigl(-\frac{k(2w-n)^2}{2n}\bigr)(1+o(1)) & \text{if $kw$ is even,}\\[\medskipamount]
0 & \text{if $kw$ is odd,}
\end{cases}
\end{equation}
uniformly for all integers $w$ in the central region.
From (\ref{eq:binomial-gaussian}) and (\ref{eq:bknw-gaussian}) we
have, 
\[
\binom{n}{w}\binom{kn}{kw}^{-1}B_k(n,w)=
\begin{cases}
\frac{2\sqrt{2}}{\sqrt{\pi n}}\exp\bigl(-\frac{(2w-n)^2}{2n}\bigr)(1+o(1)) & \text{if $kw$ is even,}\\[\medskipamount]
0 & \text{if $kw$ is odd.}
\end{cases}
\]
Thus, for $w$ in the central region,
\[
\begin{split}
\sum_{w}\binom{n}{w}\binom{kn}{kw}^{-1}B_k(n,w)&=
\bigl(1+o(1)\bigr)\frac{2\sqrt{2}}{\sqrt{\pi n}}
\sum_{w}\exp\biggl(-\frac{(2w-n)^2}{2n}\biggr)\\
&=
\bigl(1+o(1)\bigr)\frac{2\sqrt{2}}{\sqrt{\pi}}
\sum_{t}\frac{1}{\sqrt n}
\exp\biggl(-2\biggl(\frac{t}{\sqrt{n}}\biggr)^2\biggr)\\
&\longrightarrow
\begin{cases}
\frac{2\sqrt{2}}{\sqrt{\pi}}
\int_{-\infty}^{\infty}\exp\bigl(-2s^2\bigr)\,ds=2, & \text{if $k$ is even},%
\\[\medskipamount]
\frac{\sqrt{2}}{\sqrt{\pi}}
\int_{-\infty}^{\infty}\exp\bigl(-2s^2\bigr)\,ds=1, & \text{if $k$ is odd},%
\\
\end{cases}
\end{split}
\]
where the second equality follows from the change of variables
$t=w-n/2$, and the limit as $n\rightarrow\infty$ follows from the
observation that for $w$ in the central region, 
$t/\sqrt n$ ranges over $-n^{1/10}/2\leq t\leq n^{1/10}/2$; 
the halving when $k$ is odd is due to the terms associated with
odd $w$ being zero if $k$ is odd.
\end{Proof}

\begin{Cla}
\label{cla:leftlarge}
The sum in the left and right large deviation regions is $o(1)$.
\end{Cla}
\begin{Proof}
First we use an approximate variant of the saddle point method 
(see e.g.~\cite[Chap.~VIII]{FlSe06}) to derive an upper bound 
for $B_k(n,w)$. By Cauchy's coefficient formula,
\[
B_k(n,w)=\frac{1}{2\pi i}\oint\frac{E_k(z)^n}{z^{kw+1}}\,dz,
\]
where the integration contour can be taken to be a positively
oriented circle of radius $\xi>0$ centered at the origin of the complex
plane. Because $E_k(z)$ is a polynomial with positive coefficients,
the integrand assumes its maximum modulus on the contour at $z=\xi$.
Consequently, letting $\lambda=w/n$,
\begin{equation}
\label{eq:saddle-ub}
B_k(n,w)\leq
\frac{1}{2\pi i}\oint\frac{E_k(\xi)^n}{\xi^{kw+1}}\,dz=
\frac{E_k(\xi)^n}{\xi^{kw}}=
\biggl(\frac{E_k(\xi)}{\xi^{k\lambda}}\biggr)^n.
\end{equation}
As an approximation to a saddle point contour, 
let $\xi=(\lambda/(1-\lambda))^{(k-1)/k}$ and observe that 
\begin{equation}
\label{eq:xi-trick}
\exp\bigl(-(k-1)H(\lambda)\bigr)=
\frac{\xi^{\lambda k}}{\bigl(1+\xi^{k/(k-1)}\bigr)^{k-1}}.
\end{equation}
Combining
(\ref{eq:binom-bounds}), (\ref{eq:saddle-ub}) and (\ref{eq:xi-trick}) 
we have
\begin{equation}
\label{eq:largedev-bound}
\binom{n}{w}\binom{kn}{kw}^{-1}B_k(n,w)\leq
\sqrt k\biggl(\frac{E_k(\xi)}{\bigl(1+\xi^{k/(k-1)}\bigr)^{k-1}}\biggr)^n
\exp\biggl(\frac{1}{6kw}\biggr).
\end{equation}
Let $\tau^{k-1}=\xi$. 
\begin{Lem}
\label{lem:tau}
For all $\tau>0$ it holds that
$E_k(\tau^{k-1})\leq \bigl(1+\tau^k\bigr)^{k-1}$,
with equality if and only if $\tau=1$.
\end{Lem}

\begin{Proof}
Recalling (\ref{eq:ek}) and using the binomial theorem,
the inequality $E_k(\tau^{k-1})\leq \bigl(1+\tau^k\bigr)^{k-1}$ 
is easily seen to be equivalent to
\begin{equation}
\label{eq:tau-ineq1}
\sum_{j=0}^{\lfloor k/2\rfloor}\binom{k}{2j}\tau^{2j(k-1)}\leq
\sum_{j=0}^{k-1}\binom{k-1}{j}\tau^{jk}.
\end{equation}
In what follows we assume that $e$ is an even nonnegative integer;
in particular, if $e$ is used as the index of summation, then it is assumed
that $e$ runs over all even nonnegative integers.
Recalling that $\binom{k}{j}=\binom{k-1}{j}+\binom{k-1}{j-1}$
for all nonnegative integers $k$ and $j$,
it is straightforward to check that (\ref{eq:tau-ineq1}) is equivalent
to
\begin{equation}
\label{eq:tau-ineq2}
\sum_e\biggl(\binom{k-1}{e}\tau^{e(k-1)}+\binom{k-1}{e-1}\tau^{e(k-1)}\biggr)
\leq
\sum_e\biggl(\binom{k-1}{e}\tau^{ek}+\binom{k-1}{e-1}\tau^{(e-1)k}\biggr).
\end{equation}
The $e=0$ terms cancel in (\ref{eq:tau-ineq2}), so we may assume $e>0$.
To establish (\ref{eq:tau-ineq2}), we show that
\begin{equation}
\label{eq:tau-ineq3}
\binom{k-1}{e}\bigl(\tau^{ek}-\tau^{e(k-1)}\bigr)+
\binom{k-1}{e-1}\bigl(\tau^{(e-1)k}-\tau^{e(k-1)}\bigr)
\geq 0
\end{equation}
holds for each $e>0$, with equality if and only if $\tau=1$.
To this end, divide both sides of (\ref{eq:tau-ineq3})
by $\binom{k-1}{e-1}\tau^{k(e-1)}/e$ to obtain 
\begin{equation}
\label{eq:tau-ineq4}
f(\tau)=(k-e)\tau^k-k\tau^{k-e}+e\geq 0.
\end{equation}
Now observe that for $e>0$ we have $f(0)=e>0$, 
$f(1)=0$, and $f(\infty)=\infty$.
Taking the derivative of $f$, the real zeroes of
$f'(\tau)=k(k-e)\tau^{k-e-1}(\tau^e-1)$ are $-1$, $0$, and $1$.
Thus, for $\tau>0$ we have $f(\tau)\geq 0$, with equality 
if and only if $\tau=1$.
\end{Proof}

We now continue the proof of Claim \ref{cla:leftlarge}.
Observe that $\lambda\in(0,1)$ implies $\xi,\tau\in(0,\infty)$,
with $\lambda=1/2$ if and only if $\xi=\tau=1$.
Thus, for $w$ in the large deviation regions, 
that is, for $\lambda=w/n$ with $n^{-2/5}/2<|\lambda-1/2|\leq (k-1)/(2k)$,
we have
$E_k(\xi)/\bigl(1+\xi^{k/(k-1)}\bigr)^{k-1}<1$ 
in (\ref{eq:largedev-bound}) by Lemma \ref{lem:tau}.
Developing $E_k(\xi)/\bigl(1+\xi^{k/(k-1)}\bigr)^{k-1}$ 
into a truncated Taylor series at $\lambda=1/2$ and evaluating
at $\lambda=(1\pm n^{-2/5})/2$, we obtain
\begin{equation}
\label{eq:largedev-bound2}
\biggl(\frac{E_k(\xi)}{\bigl(1+\xi^{k/(k-1)}\bigr)^{k-1}}\biggr)^n
\leq \exp\biggl(-\frac{(k-1)n^{1/5}}{2k}\biggr)\bigl(1+o(1)\bigr)
\end{equation}
uniformly for all $w$ in the left and right large deviation regions.
The claim now follows from (\ref{eq:largedev-bound}) and 
(\ref{eq:largedev-bound2}) because the regions have $O(n)$ summands.
\end{Proof}

\begin{Cla}
The sum in the left extreme deviation region is $1+o(1)$.
\end{Cla}
\begin{Proof}
For $w=0$ the term in the sum is $1$.
For $w=1,2$ the terms are $O(n^wn^{-kw}n^{kw/2})$.
Thus, in what follows we may restrict to $3\leq w<n/(2k)$. 
Observe that
\begin{equation}
\label{eq:bknw-roughbound}
B_k(n,w)\leq \binom{kw/2+n-1}{n-1}\binom{k}{2}^{kw/2}.
\end{equation}
Indeed, $\binom{kw/2+n-1}{n-1}$ counts the number of integer
compositions of $kw$ into $n$ even nonnegative parts, and 
$\binom{k}{2}^{kw/2}$ provides an upper bound for the product
$\prod_{i=1}^n\binom{k}{e_i}$ associated with each 
composition $e_1+e_2+\ldots+e_n=kw$ into even nonnegative parts
at most $k$.

Observe by (\ref{eq:binom-bounds}) and (\ref{eq:bknw-roughbound}) that
\[
\log\biggl(\binom{n}{w}\binom{kn}{kw}^{-1}B_k(n,w)\biggr)\leq G_k(n,w)+O(1),
\]
where
\[
G_k(n,w)=
-(k-1)n H\biggl(\frac{w}{n}\biggr)+
\biggl(\frac{kw}{2}+n-1\biggr)H\biggl(\frac{n-1}{kw/2+n-1}\biggr)+
\frac{kw}{2}\log\binom{k}{2}.
\]
Differentiating twice with respect to $w$, we have
\[
G_k''(n,w)=\frac{k(kn-1)w+n(n-1)(k-2)}{(n-w)w(kw+2n-2)}
\]
is positive for $0<w<n$, implying that $G_k$ is convex 
in the extreme deviation region. In particular, 
$G_k$ assumes its maximum at the boundaries of the region. Evaluating 
$G_k$ at $w=3$ and $w=n/(2k)$, we find 
$G_k(n,w)\leq -3/2\log(n)+O(1)$ uniformly for all $w$ in the region.
The claim follows because the number of terms in the region is $O(n)$.
\end{Proof}

Combining the results for all the regions, we have
that $S_k(n)=2+o(1)$ if $k$ is odd and $S_k(n)=4+o(1)$ if $k$ is even.
This completes the proof of Theorem \ref{thm:sk}.
\end{Proof}

\section{Topographical Properties}
\label{sect:topo}

Throughout this section we consider
the landscape associated with a system $Ax\equiv 0\pmod 2$,
where $A$ is a $k$-regular matrix of size $n\times n$ 
that both (a) has a kernel of size at most 
$2^d$ and (b) is a $(k,\beta n,k-2-\delta)$-boundary expander,
where $0<\beta<1/2$, $d>0$, and $0<\delta<1/3$ are constants independent 
of $n$. 

\subsection{Energy Barriers and Local Minima}
\label{sect:main-sketch}

The intuition underlying Theorem \ref{thm:main} is as follows.
The boundary expansion property in effect ``surrounds'' a ground state 
with a ``perimeter'' of radius $\lfloor\beta n\rfloor$ in the $n$-dimensional
hypercube, where the energy (``wall'') at every perimeter 
state is at least $(k-2-\delta)\lfloor\beta n\rfloor$, so any state 
outside the perimeter with considerably lower energy has a 
considerable barrier separating it from the ground state.

Let us now make this intuition formally precise and prove 
Theorem \ref{thm:main}.

Property (i) is immediate by assumption.
To establish property (ii), let $g_1$ and $g_2$ be any two
distinct ground states.
Clearly, $D(g_1,g_2)=W(g_1+g_2)>0$ and $Ag_1=Ag_2=0$.
Thus, it follows from the boundary expansion property that 
$D(g_1,g_2)=W(g_1+g_2)>\beta n$. 
(Indeed, we cannot have $0=W(A(g_1+g_2))\geq (k-2-\delta)W(g_1+g_2)>0$.)
Thus, any walk of successive adjacent states from $g_1$ to $g_2$ must
have a ``perimeter'' state $p$ with 
$D(g_1,p)=W(g_1+p)=\lfloor \beta n\rfloor$.
By the boundary expansion property,
\[
E(p)-E(g_1)=W(Ap)-W(Ag_1)=W(Ap)=W(A(g_1+p))\geq (k-2-\delta)W(g_1+p)=
(k-2-\delta)\lfloor \beta n\rfloor.
\]
Since $g_1$ and $g_2$ were arbitrary, we have thus established that 
distinct ground states are at distance $\Omega(n)$ and separated by
an $\Omega(n)$ energy barrier.

To establish property (iii),
we first require a large enough set of local minima.
With foresight, select any constant $\gamma$ such that 
\[
0<\gamma<\min\biggl(\frac{\beta(k-2-\delta)}{4}\,,\,
                    \frac{1}{2^d(k(k-1)+1)}\biggr).
\]
Because the kernel of $A$ has dimension at most $d$,
by elementary linear algebra there is a linearly independent 
set of $n-d$ columns of $A$.
Furthermore, $A$ restricted to these columns has a linearly
independent set of $n-d$ rows. 
By permuting the rows if necessary, we can assume that these rows 
occur first in $A$. 
Applying Gaussian elimination to the selected $n-d$ linearly 
independent columns, we find $n-d$ vectors $y_1,y_2,\ldots,y_{n-d}$ 
with the property that $Ay_j=e_j+r_j$, where 
$r_j$ is a vector with the first $n-d$ entries equal to $0$, 
and $e_j$ is the $j$th vector in the standard basis.
Observe that the vectors $y_1,y_2,\ldots,y_{n-d}$ are linearly 
independent.

We say that $y_j$ \dtext{marks} the rows that contain a 1 in $A$
in at least one of the columns containing a 1 in row $j$. 
In other words, denoting by $a_{pq}$ the entry of $A$ at row $p$, column $q$,
we have that $y_j$ marks the rows $\{i:\exists\,q\ a_{iq}=a_{jq}=1\}$.
Observe that because $A$ is $k$-regular, each vector $y_j$ marks 
at most $k(k-1)+1$ rows. 

There are at most $2^d$ different vectors $r_j$. Thus, 
because $d$ is a fixed constant independent of $n$, 
there exist at least $(n-d)/2^d$ vectors $y_j$ that have identical 
associated vectors $r_j$.
Among these vectors, start selecting vectors one by one 
and marking associated rows subject to the constraint
that no row is marked more than once, until no more vectors can be selected.
Let $m$ be the number of vectors selected in this way.
Clearly, $(n-d)/(2^d(k(k-1)+1))\leq m\leq n$.  
By re-indexing the vectors and permuting the rows and columns of $A$ 
if necessary, we can assume that the selected vectors are
$y_1,y_2,\ldots,y_m$.

We now claim that every state $u$ of the form 
$u=\sum_{j=1}^m\chi_j y_j$ with $\chi_j\in\{0,1\}$ and 
$\sum_{j=1}^m\chi_j\equiv 0\pmod 2$ is a local minimum.
To see this, observe first that $Au=\sum_{j=1}^m\chi_j e_j$
and that $E(u)=\sum_{j=1}^m\chi_j$.
Now the marking constraint implies that if we flip the value of
any one variable in $u$, we satisfy at most one violated equation
and introduce at least $k-1$ new violated equations. Thus,
$u$ is a local minimum.

We proceed to construct an auxiliary graph that we eventually
use to establish the energy barriers separating certain
local minima from all the ground states.
For $j=1,2,\ldots,m-1$, let $z_j=y_j+y_m$.
Observe that the vectors $z_1,z_2,\ldots,z_{m-1}$ are linearly
independent. Furthermore, the $\binom{m-1}{2}$ 
sums of the form $z_i+z_j$ with 
$1\leq i<j\leq m-1$ clearly satisfy $W(A(z_i+z_j))=2$.

Because $A$ is $k$-regular, 
there are at most $\binom{k}{2}n=O(n)=O(m)$ vectors 
$y$ with $W(Ay)=2$ and $W(y)\leq 2/(k-2-\delta) < 3$.
To see this, observe that the two columns selected by
any vector $y$ with $W(y)=2$ and $W(Ay)=2$ must have at least
one row containing a $1$ in both columns, and the total 
number of such ``$11$''-patterns in $A$ is $\binom{k}{2}n$. 
Thus, by the expansion property $\binom{m-1}{2}-O(m)$ of 
the sums $z_i+z_j$ satisfy $W(z_i+z_j)> \beta n$.

Form an auxiliary graph with the vertex set $\{z_1,z_2,\ldots,z_{m-1}\}$ 
such that any two distinct vertices, $z_i$ and $z_j$, are adjacent 
if and only if $W(z_i+z_j)\leq \beta n$. 
Because the number of edges in the auxiliary graph is $O(m)$, 
for all sufficiently large $n$ the auxiliary graph has an independent 
set of size $2^d+1$, which---by relabeling if necessary---can be 
assumed to consist of the vectors $z_1,z_2,\ldots,z_{2^d+1}$.

We now construct the local minima meeting property (iii).
Let $g_1,g_2,\ldots,g_s$ be the ground states, $s\leq 2^d$. 
Observe that any sum consisting of a subset of 
the linearly independent vectors $z_1,z_2,\ldots,z_{m-1}$ is a 
local minimum. Furthermore, the energy of such a minimum is at most 
the number of summands plus one.
Select any $\lceil\gamma n\rceil$ of the vectors 
$z_{2^d+2},z_{2^d+3},\ldots,z_{m-1}$.
(Note that for all sufficiently large $n$ this is possible due to 
the choice of $\gamma$.)

Consider now any state $u$ formed as the sum of a nonempty subset of 
the $\lceil\gamma n\rceil$ selected vectors. 
The energy of $u$ is $E(u)=W(Au)\leq \lceil\gamma n\rceil+1$.
The state $u$ does not necessarily have extensive barriers separating
it from each of the ground states. However,
if the following condition holds, then $u$ is separated by 
extensive barriers from the ground states. 
If the condition does not hold, then
adding one of the (independent) vectors $z_1,z_2,\ldots,z_{2^d+1}$ 
to $u$ will produce a local minimum that is separated by 
extensive barriers from the ground states.

Suppose that $D(u,g_j)>\beta n/2$ holds for all $j=1,2,\ldots,s$.
Thus, for every solution state $g_j$, any walk from $u$ to $g_j$
consisting of successive adjacent states must 
contain a ``perimeter'' state $p$ at distance 
$D(p,g_j)=W(p+g_j)=\lceil \beta n/2\rceil$
By the boundary expansion property,
the energy of $p$ is 
\[
E(p)=W(Ap)=W(Ap+Ag_j)=W(A(p+g_j))\geq
(k-2-\delta) W(p+g_j)\geq\frac{(k-2-\delta)\beta n}{2}.
\]
In particular, the increase in energy at $p$ compared with
the energy of $u$ is
\[
E(p)-E(u)=
W(Ap)-W(Au)\geq
\frac{(k-2-\delta)\beta n}{2}-\lceil\gamma n\rceil-1>
\gamma n.
\]
Thus, the energy barrier separating $u$ 
from $g_j$ is at least $\gamma n$,
assuming that $D(u,g_j)>\beta n/2$ holds for all $j=1,2,\ldots,s$.

Suppose that $D(u,g_j)\leq \beta n/2$ holds for at least one 
$j=1,2,\ldots,s$. Then, we claim that there exists 
at least one $\ell=1,2,\ldots,2^d+1$ such that
$D(u+z_\ell,g_j)>\beta n/2$ holds for all $j=1,2,\ldots,s$.
To reach a contradiction, suppose that this is not the case.
Then, by the pigeonhole principle, there exists
a $j=1,2,\ldots,s$ and $1\leq\ell_1<\ell_2\leq 2^d+1$
such that $D(u+z_{\ell_1},g_j)\leq\beta n/2$ and
$D(u+z_{\ell_2},g_j)\leq\beta n/2$.
By the triangle inequality, 
$D(u+z_{\ell_1},u+z_{\ell_2})\leq \beta n$, 
which by $D(z_{\ell_1},z_{\ell_2})=D(u+z_{\ell_1},u+z_{\ell_2})$
contradicts the fact that $z_1,z_2,\ldots,z_{2^d+1}$ form an independent
set in the auxiliary graph.
Therefore, there exists at least one $\ell=1,2,\ldots,2^d+1$
such that $D(u+z_\ell,g_j)>\beta n/2$ holds for all $j=1,2,\ldots,s$.
Applying the argument in the previous paragraph to the state $u+z_\ell$,
we have that $u+z_\ell$ is separated from every solution state 
by an energy barrier at least $\gamma n$.

Because the vectors $z_1,z_2,\ldots,z_{m-1}$ are linearly independent,
we have thus established the existence of at least $2^{\gamma n}-1$ 
distinct local minima, each separated from every ground state by an 
energy barrier of at least $\gamma n$. This establishes property (iii).

\subsection{Lower Bound for the Focused Random Walk}

The intuition underlying Theorem \ref{thm:main2} is as follows.
Consider any ground state $g$.
In any state $s\neq g$ with $D(s,g)\leq\lfloor\beta n\rfloor$, 
the boundary expansion 
property implies that most equations violated by $s$ in the system 
$Ax\equiv 0\pmod 2$ have \emph{exactly one} variable that assumes 
different values in $s$ and $g$. In particular, the focused
random walk is unlikely to flip this variable 
(there are $k-1$ other choices), thereby exerting a systematic 
drift away from $g$. Thus, expansion in effect induces a region 
of entropic repulsion around every ground state.

Let us now make this intuition formally precise and prove 
Theorem \ref{thm:main2}. As was demonstrated in \S\ref{sect:main-sketch}, 
all ground states in the landscape have distance at least 
$\lceil\beta n\rceil$.
Because $\beta<1/2$, it follows by standard tail bounds for the
binomial distribution (see e.g.~\cite[\S 1]{Boll01}) that
with probability $1-2^{-\Omega(n)}$ the random initial state for 
the focused random walk has distance at least $\lceil\beta n\rceil$ 
to each of the at most $2^d$ ground states. 
It thus suffices to show that the expected number of steps 
to reach the ground state from the perimeter of a region of repulsion 
is $2^{\Omega(n)}$. To this end, let $g$ be any ground state, and
consider any state $s\neq g$ with 
$D(s,g)=W(s+g)\leq\lfloor\beta n\rfloor$. 
Because $Ag=0$, the number of violated
equations in $s$ is $E(s)=W(As)=W(A(s+g))=E(s+g)$.
By $k$-regularity, $E(s+g)\leq k W(s+g)$. By the expansion property
and $As=A(s+g)$, at least $(k-2-\delta)W(s+g)$ equations violated 
by $s$ contain exactly one variable having a different value in 
$s$ and $g$. Thus, assuming that the focused random walk is in 
the state $s$, the random step will increase the distance to $g$ 
by $1$ with probability at least
\begin{equation}
\label{eq:drift}
\frac{(k-1)(k-2-\delta)W(s+g)}{kE(s)}
\geq
\frac{(k-1)(k-2-\delta)W(s+g)}{k^2W(s+g)}\geq
\frac{(k-1)(k-2-\delta)}{k^2},
\end{equation}
otherwise the distance to $g$ decreases by $1$.
For $k\geq 6$ the probability (\ref{eq:drift}) is at least $55/108$, 
thereby establishing a systematic drift away from $g$ for 
every state $s\neq g$ with $D(s,g)\leq\lfloor\beta n\rfloor$. 
A standard analysis of the gambler's ruin problem 
(see e.g.~\cite[Chap.~XIV]{Fell57}) with one absorbing ruin state 
and one reflecting barrier now establishes that, 
starting from a (reflecting) state $s$ at distance
$D(s,g)\geq\beta n$ from each ground state $g$, 
the expected number of steps required to reach a ground
(ruin) state is $2^{\Omega(n)}$.



\clearpage
\thispagestyle{empty}

\renewcommand{\thepage}{\textsc{Appendix~p.~\arabic{page}}}
\appendix
\setcounter{page}{1}
\noindent
\begin{center}
{\large\textsc{Appendix}}
\end{center}

This appendix is provided only for convenience of verification of the
earlier results. In particular, we stress that Theorems 
\ref{thm:expansion} and \ref{thm:local-limit} are well known; 
cf.~\cite[Theorem~4.16(2)]{HoLW06} and 
\cite{Haym56}, \cite[Chaps.~VIII and IX]{FlSe06}.

\section{Proof of Theorem \ref{thm:expansion}}
\label{sect:expansion-proof}

A matrix $A$ is a $(k,\omega,\eta)$-\dtext{expander} 
if
(a) the number of nonzero entries in every column is at most $k$,
and 
(b) for all $w=1,2,\ldots,\lfloor\omega\rfloor$, every submatrix 
    consisting of $w$ columns of $A$ has at least $\lceil\eta w\rceil$ rows 
    containing at least one nonzero value.

Theorem \ref{thm:expansion} follows immediately by combining the following 
two results.

\begin{Lem}
\label{lem:boundary-expansion}
Let $A$ be a $(k,\omega,\eta)$-expander.
Then, $A$ is a $(k,\omega,2\eta-k)$-boundary expander.
\end{Lem}
\begin{Proof}
Consider any submatrix of $A$ consisting of $w$ of its columns, 
$1\leq w\leq \omega$. Denote by $c_\ell$ the number of rows with 
exactly $\ell$ nonzero values in these columns. Because $A$ is an 
expander, we have
$C=c_1+c_2+\ldots+c_w\geq \eta w$ and $C'=c_1+2c_2+\ldots+wc_w\leq kw$.
In particular, $c_1\geq 2C-C'\geq (2\eta-k)w$.
\end{Proof}

\begin{Thm}
For every\/ $\delta>0$ there exists an $\beta>0$ 
such that a random $k$-regular matrix
is a\/ $(k,\beta n,k-1-\delta)$-expander with probability\/ $1-o(1)$. 
\end{Thm}
\begin{Proof}
Select a $0<\delta<1$. Let $\eta=k-1-\delta$ and
\begin{equation}
\label{eq:u-def}
\begin{split}
U_k(n,w)=\binom{n}{w}
         \binom{n}{\lfloor \eta w \rfloor}
         \frac{(k\lfloor \eta w \rfloor)!}{(k\lfloor \eta w \rfloor-kw)!}
         \frac{(kn-kw)!}{(kn)!}
        =\binom{n}{w}
         \binom{n}{\lfloor \eta w \rfloor}
         \binom{k\lfloor \eta w \rfloor}{kw}
         \binom{kn}{kw}^{-1}.
\end{split}
\end{equation}
Recalling the configuration model from \S\ref{sect:configuration},
we claim that the probability that a random $(k,n)$-configuration 
does not define a $(k,\beta n,\eta )$-expander is bounded from above by 
$\sum_{w=1}^{\lfloor \beta n\rfloor} U_k(n,w)$.
To see this, observe that for every configuration violating
the expansion property, there exists a set of $w$ cells in $\mathcal{J}$ 
and a set of $\lfloor \eta w\rfloor$ cells in $\mathcal{I}$ such that 
the $kw$ points in the former set of cells are paired with points 
in the latter.
There are $(k\lfloor \eta w \rfloor)!/(k\lfloor \eta w \rfloor-kt)!$ ways 
to pair the $kw$ points, and $(kn-kw)!$ ways to pair the remaining points.

We proceed to show that 
$\sum_{w=1}^{\lfloor \beta n\rfloor} U_k(n,w)=o(1)$ 
for an appropriate fixed $\beta>0$.
For $w=1,2,\ldots,\lfloor 2/\delta\rfloor$
we may view $U_k(n,w)$ as a rational function of two polynomials 
of $n$. The denominator polynomial has degree $kw$ and the 
numerator polynomial has degree 
$w+\lfloor \eta w\rfloor=\lfloor (k-\delta)w\rfloor\leq kw-1$.
Thus, $\sum_{w=1}^{\lfloor 2/\delta\rfloor}U_k(n,w)=O(1/n)$.
Now let
\begin{equation}
\label{eq:lk}
L_k(n,w)=n H\biggl(\frac{\eta w}{n}\biggr)+
         k\eta wH\biggl(\frac{1}{\eta}\biggr)-
         (k-1)n H\biggl(\frac{w}{n}\biggr)
\end{equation}
and observe by (\ref{eq:binom-bounds}) and (\ref{eq:u-def})
that $\log U_k(n,w)\leq L_k(n,w)+O(1)$ for all large enough $n$ and
$w=\lfloor 2/\delta\rfloor+1,
   \lfloor 2/\delta\rfloor+2,\ldots,\lfloor n/(2\eta)\rfloor$.
Observe that $L_k(n,2/\delta)\leq -2\log(n)+O(1)$ and that,
differentiating (\ref{eq:lk}) with respect to $w$ and letting $\lambda=w/n$, 
\begin{equation}
\label{eq:lkd}
L_k'(n,w)=k\eta H\biggl(\frac{1}{\eta}\biggr)+
\log\biggl(\frac{\lambda^\delta(1-\eta\lambda)^\eta}{\eta^\eta(1-\lambda)^{k-1}}\biggr).
\end{equation}
Because the log-term in (\ref{eq:lkd}) decreases without bound 
as $\lambda\rightarrow 0^+$,
there exists an $\beta>0$ such that $L_k(n,w)$ is decreasing 
as $w=\lfloor 2/\delta\rfloor+1,
      \lfloor 2/\delta\rfloor+2,
      \ldots,\lfloor\beta n\rfloor$.
Thus, 
$\sum_{w=1}^{\lfloor\beta n\rfloor}U_k(n,w)<
 O(1/n)+\beta n/(n^2\cdot O(1))=O(1/n)$.
The claim now follows from Theorem \ref{thm:simple}.
\end{Proof}

\section{Proof of Theorem \ref{thm:local-limit}}
\label{sect:gaussian-proof}

We require first a preliminary result. 

\begin{Thm}[Saddle point asymptotics for coefficients of a polynomial power]
\label{thm:saddle}
Let $P(z)$ be a polynomial of degree $d\geq 1$ with 
a positive constant term and positive coefficients
such that the greatest common divisor of the degrees of the 
nonzero terms of $P(z)$ is\/ $1$, 
and let $\Lambda$ be any compact 
subinterval of the open interval $(0,d)$.
Then, for all large enough $n$, it holds uniformly 
for all integers $N=\lambda n$ with $\lambda\in\Lambda$ 
that
\begin{equation}
\label{eq:saddlepower}
[z^N]\bigl\{P(z)^n\bigr\}=
\frac{1}{\sqrt{2\pi n K_\lambda''(\xi)}}
\frac{P(\xi)^n}{\xi^{\lambda n+1}}(1+o(1)),
\end{equation}
where $K_\lambda(z)=\log(P(z))-\lambda\log(z)$ and
$\xi=\xi(\lambda)$ is the unique positive solution 
of $K_\lambda'(\xi)=0$.
\end{Thm}
\begin{Proof}
It follows from the assumptions on $P$ that $\xi$ exists and is unique
for every $\lambda\in\Lambda$.
Applying Cauchy's coefficient formula on the circular contour
$z(\theta)=\xi\exp(i\theta)$, $-\pi\leq\theta\leq\pi$, we
have
\[
[z^N]\bigl\{P(z)^n\bigr\}=
\frac{1}{2\pi i}\int_{-\pi}^\pi\frac{P(\xi\exp(i\theta))^n}{(\xi\exp(i\theta))^{N+1}}\xi i\exp(i\theta)\,d\theta=
\frac{1}{2\pi}\int_{-\pi}^\pi\frac{P(\xi\exp(i\theta))^n}{(\xi\exp(i\theta))^{\lambda n}}\,d\theta.
\]
By the assumptions on $P$, the modulus of $P(z)$ assumes its maximum 
value on the contour if and only if $z=\xi$.
Thus, as $n$ increases, the neighborhood of $\xi$ produces 
the (exponentially) dominant contribution to the integral. 
In particular, letting $\theta_0=n^{-2/5}$, we have
\[
[z^N]\bigl\{P(z)^n\bigr\}=
\frac{1}{2\pi}(1+o(1))\int_{-\theta_0}^{\theta_0}
\frac{P(\xi\exp(i\theta))^n}{(\xi\exp(i\theta))^{\lambda n}}\,d\theta=
\frac{1}{2\pi}(1+o(1))\int_{-\theta_0}^{\theta_0}
\exp(nK_\lambda(\xi\exp(i\theta)))\,d\theta.
\]
Assuming that $n$ is large enough, $K_\lambda$ is analytic on every 
straight line segment connecting
$\xi$ to $\xi\exp(i\theta)$ with $-\theta_0\leq\theta\leq\theta_0$.
Thus, we have (see e.g.~\cite[\S7.1]{WhWa63}), 
\[
\begin{split}
K_\lambda(\xi\exp(i\theta))&=
K_\lambda(\xi)+
K_\lambda'(\xi)(\xi\exp(i\theta)-\xi)+
K_\lambda''(\xi)(\xi\exp(i\theta)-\xi)^2/2+\\
&\qquad\frac{(\xi\exp(i\theta)-\xi)^3}{2}\int_0^1(t-1)^2K_\lambda'''(\xi+t(\xi\exp(i\theta)-\xi))\,dt.
\end{split}
\]
By assumption we have that $K'(\xi)=0$.
Furthermore, the last integral and $\xi$ are bounded when $\lambda\in\Lambda$.
Thus, using $\exp(i\theta)=1+i\theta+O(n^{-4/5})$ for the
second-order term and $\exp(i\theta)=1+O(n^{-2/5})$ for the coefficient
of the integral, we have 
\[
K_\lambda(\xi\exp(i\theta))=
K_\lambda(\xi)-K_\lambda''(\xi)\xi^2\theta^2/2+
O(n^{-6/5})
\]
uniformly for $-\theta_0\leq\theta\leq\theta_0$.
It follows that
\[
\begin{split}
[z^N]\bigl\{P(z)^n\bigr\}&=
\frac{1}{2\pi}(1+o(1))\frac{P(\xi)^n}{\xi^{\lambda n}}%
\int_{-\theta_0}^{\theta_0}\exp(-nK_\lambda''(\xi)\xi^2\theta^2/2)\,d\theta
\\
&=\frac{1}{2\pi\sqrt n}(1+o(1))\frac{P(\xi)^n}{\xi^{\lambda n}}%
\int_{-\theta_0\sqrt n}^{\theta_0\sqrt n}\exp(-K_\lambda''(\xi)\xi^2t^2/2)\,dt
\\
&=\frac{1}{2\pi\sqrt n}(1+o(1))\frac{P(\xi)^n}{\xi^{\lambda n}}%
\int_{-\infty}^{\infty}\exp(-K_\lambda''(\xi)\xi^2t^2/2)\,dt
\\
&=\frac{1}{\sqrt{2\pi n K_\lambda''(\xi)}}\frac{P(\xi)^n}{\xi^{\lambda n+1}}(1+o(1)),
\end{split}
\]
where the second equality follows from the change of variables 
$\theta=t/\sqrt{n}$ and the last equality follows from 
$\int_{-\infty}^{\infty}\exp(-t^2)\,dt=\sqrt{\pi}$.
\end{Proof}

We now proceed with the proof of Theorem \ref{thm:local-limit}.

\begin{Proof}
Observe that $0<\mu<d$ and let $\Lambda$ be any compact subinterval
of $(0,d)$ with $\mu$ in its interior.
We proceed to apply Theorem \ref{thm:saddle} with sufficient approximations
to $\xi$, $K_\lambda''(\xi)$, and $P(\xi)/\xi^\lambda$ as 
$n$ increases and $\lambda=N/n=\mu+\nu/n$ with 
$|\lambda-\mu|\leq n^{\delta-1}$.
First, observe that $\lambda=\mu$ implies $\xi(\lambda)=1$.
Developing $\xi(\lambda)$ into a truncated Taylor series at $\lambda=\mu$
using the defining equality 
$\xi(\lambda) P'(\xi(\lambda))=\lambda P(\xi(\lambda))$,
we have, after some calculation, uniformly
\begin{equation}
\label{eq:gauss-apx1}
\xi(\lambda)
=1+\xi'(\mu)(\lambda-\mu)+O\bigl(n^{2(\delta-1)}\bigr)
=1+\frac{\nu}{\sigma^2n}+O\bigl(n^{2(\delta-1)}\bigr).
\end{equation}
Some more calculation gives the uniform approximations
\begin{equation}
\label{eq:gauss-apx2}
K_\lambda''(\xi(\lambda))
=\frac{\lambda}{\xi(\lambda)^2}-\frac{P'(\xi(\lambda))^2}{P(\xi(\lambda))^2}+
 \frac{P''(\xi(\lambda))}{P(\xi(\lambda))}=
\sigma^2+O(n^{\delta-1})
\end{equation}
and
\begin{equation}
\label{eq:gauss-apx3}
\frac{P(\xi(\lambda))}{\xi(\lambda)^\lambda}=
P(1)-\frac{P(1)}{2\sigma^2}(\lambda-\mu)^2+O\bigl(n^{3(\delta-1)}\bigr)=P(1)\biggl(1+\frac{-\nu^2/(2\sigma^2n)+O(n^{3\delta-2})}{n}\biggr).
\end{equation}
The approximations (\ref{eq:gauss-apx1}), (\ref{eq:gauss-apx2}), 
and (\ref{eq:gauss-apx3}) applied to (\ref{eq:saddlepower}) 
establish (\ref{eq:gaussianpower}).
\end{Proof}


\begin{thebibliography}{XX}

\bibitem{AaKo89}
E.~Aarts, J.~Korst,
\emph{Simulated Annealing and Boltzmann Machines},
Wiley, Chichester, 1989.

\bibitem{AaLe97}
E.~Aarts, J.K.~Lenstra,
\emph{Local Search in Combinatorial Optimization},
Wiley, Chichester, 1997.

\bibitem{AcNP05}
D.~Achlioptas, A.~Naor, Y.~Peres,
Rigorous locations of phase transitions in hard optimization
problems,
\emph{Nature} 435 (2005) 759--764.

\bibitem{AcPe04}
D.~Achlioptas, Y.~Peres,
The threshold for random $k$-SAT is $2^k\log 2-O(k)$,
\emph{J. Amer.\ Math.\ Soc.}\ 17 (2004) 947--973.

\bibitem{AcRi06}
D.~Achlioptas, F.~Ricci-Tersenghi,
On the solution-space geometry of random constraint
satisfaction problems,
in: Proc.~28th ACM Symposium on Theory of Computing 
(Seattle, May 21--23, 2006), ACM Press, New York, 2006, pp.~130--139.

\bibitem{AlBS03}
M.~Alekhnovich, E.~Ben-Sasson,
Linear upper bounds for random walk on small density random 3-CNFs,
in: Proc.~44th IEEE Symposium on Foundations of Computer Science
(Cambridge, Mass., Oct.~11--14, 2003),
IEEE Computer Society Press, Los Alamitos, Calif., 2003, 
pp.~352--361.

\bibitem{ABRW04}
M.~Alekhnovich, E.~Ben-Sasson, A.A.~Razborov, A.~Wigderson,
Pseudorandom generators in propositional proof complexity,
\emph{SIAM J. Comput.}\ 34 (2004) 67--88.

\bibitem{ABBI05}
M.~Alekhnovich, A.~Borodin, J.~Buresh-Oppenheim, R.~Impagliazzo,
A.~Magen, T.~Pitassi,
Toward a model for backtracking and dynamic programming,
in: Proc. 12th IEEE Conference on Computational Complexity
(San Jose, Calif., June 11--15, 2005),
IEEE Computer Society Press, Los Alamitos, Calif., 2005, pp.~308--322.

\bibitem{AlHI05}
M.~Alekhnovich, E.A.~Hirsch, D.~Itsykson,
Exponential lower bounds for the running time of DPLL algorithms
on satisfiable formulas,
\emph{J. Automat.\ Reason.}\ 35 (2005) 51--72.

\bibitem{ArAu06}
J.~Ardelius, E.~Aurell,
Behavior of heuristics on large and hard satisfiability problems,
\emph{Phys.\ Rev.\ E} 74 (2006) 037702.

\bibitem{AuGK05}
E.~Aurell, U.~Gordon, S.~Kirkpatrick,
Comparing beliefs, surveys and random walks,
in: L.K.~Saul, Y.~Weiss, L.~Bottou (Eds.),
\emph{Advances in Neural Information Processing Systems\/ $17$},
MIT Press, Cambridge, Mass., 2005, pp.~49--56.

\bibitem{BaBu05}
O.~Barak, D.~Burshtein,
Lower bounds on the spectrum and error rate of LDPC code ensembles,
in: Proc. 2005 IEEE Symposium on Information Theory 
(Adelaide, Sept.~4--9, 2005), 
IEEE, New York, 2005, pp.~42--46.

\bibitem{BaHW03}
W.~Barthel, A.K.~Hartmann, M.~Weigt,
Solving satisfiability problems by fluctuations: 
the dynamics of stochastic local search,
\emph{Phys.\ Rev.\ E} 67 (2003) 066104.

\bibitem{BeBK72}
A.~B\'ek\'essy, P.~B\'ek\'essy, J. Koml\'os,
Asymptotic enumeration of regular matrices,
\emph{Studia Sci.\ Math.\ Hungar.}\ 7 (1972) 343--353.

\bibitem{BeWi01}
E.~Ben-Sasson, A.~Wigderson, 
Short proofs are narrow---resolution made simple,
\emph{J.\ ACM}\ 48 (2001) 149--169.

\bibitem{Bend73}
E.A.~Bender, 
Central and local limit theorems applied to asymptotic enumeration,
\emph{J.\ Combin.\ Theory Ser.\ A} 15 (1973) 91--111.

\bibitem{Boll01}
B.~Bollob\'as,
\emph{Random Graphs}, 2nd ed.,
Cambridge University Press, Cambridge, 2001.

\bibitem{BoBo07}
E.~Bolthausen, A.~Bovier (Eds.),
\emph{Spin Glasses},
Springer, 2007, to appear.

\bibitem{BrMZ05}
A.~Braunstein, M.~M\'ezard, R.~Zecchina,
Survey propagation: an algorithm for satisfiability,
\emph{Random Structures Algorithms} 27 (2005) 201--226.

\bibitem{BKMT06}
Z.~Burda, A.~Krzywicki, O.C.~Martin, Z.~Tabor,
From simple to complex networks: inherent structures, barriers,
and valleys in the context of spin glasses,
\emph{Phys.\ Rev.\ E} 73 (2006) 036110.

\bibitem{BuMi04}
D.~Burshtein, G.~Miller,
Asymptotic enumeration methods for analyzing LDPC codes,
\emph{IEEE Trans.\ Inform.\ Theory} 50 (2004) 1115--1131.

\bibitem{Cato99}
O.~Catoni,
Simulated annealing algorithms and Markov chains with rare transitions,
\emph{S\'eminaire de Probabilit\'es, XXXIII}, 
Springer, Berlin, 1999, pp.~69--119.

\bibitem{Cern85}
V.~\v Cern\'y,
Thermodynamical approach to the traveling salesman problem:
an efficient simulation algorithm,
\emph{J.\ Optim.\ Theory Appl.}\ 45 (1985) 41--51.

\bibitem{ChKT91}
P.~Cheeseman, B.~Kanefsky, W.M.~Taylor,
Where the really hard problems are,
in: J.~Mylopoulos, R.~Reiter (Eds.), 
Proc.~12th International Joint Conference on Artifical Intelligence
(Sydney, Aug.~24--30, 1991), 
Morgan Kaufmann, San Francisco, 1991, pp.~331--337.

\bibitem{ChSz88}
V.~Chv\'atal, E.~Szemer\'edi, 
Many hard examples for resolution,
\emph{J.\ ACM}\ 35 (1988) 759--768.

\bibitem{CoDM03}
S.~Cocco, O.~Dubois, J.~Mandler, R.~Monasson,
Rigorous decimation-based construction of ground pure states
for spin-glass models on random lattices,
\emph{Phys.\ Rev.\ Lett.}\ 90 (2003) 047205.

\bibitem{CrDa99}
N.~Creignou, H.~Daude,
Satisfiability threshold for random XOR-CNF formulas,
\emph{Discrete Appl.\ Math.}\ 96/97 (1999) 41--53.

\bibitem{DaSi03}
J.~Dall, P.~Sibani,
Exploring valleys of aging systems: the spin glass case,
\emph{Eur. Phys. J.} 36 (2003) 233--243.

\bibitem{DiRU06}
C.~Di, T.J.~Richardson, R.L.~Urbanke,
Weight distribution of low-density parity-check codes,
\emph{IEEE Trans.\ Inform.\ Theory} 52 (2006) 4839--4855.

\bibitem{DiSe05}
Z.~Dietz, S.~Sethuraman,
Large deviations for a class nonhomogeneous Markov chains,
\emph{Ann.\ Appl.\ Probab.}\ 15 (2005) 421--486.

\bibitem{DuGP97}
D.~Du, J.~Gu, P.M.~Pardalos (Eds.),
\emph{Satisfiability Problem: Theory and Applications},
American Mathematical Society, Providence, R.I., 1997.

\bibitem{DuMa02}
O.~Dubois, J.~Mandler,
The 3-XORSAT threshold,
in: Proc.~43rd IEEE Symposium on Foundations of Computer Science
(Vancouver, Nov.~16--19, 2002)
IEEE Computer Society Press, Los Alamitos, Calif., 2002, 
pp.~769--778.

\bibitem{DMSZ01}
O.~Dubois, R.~Monasson, B.~Selman, R.~Zecchina (Eds.),
Phase transitions in combinatorial problems,
\emph{Theoret.\ Comput.\ Sci.}\ 265 (2001) no.~1--2.

\bibitem{ErKa46}
P.~Erd\"os, I.~Kaplansky,
The asymptotic number of Latin rectangles,
\emph{Amer.\ J.\ Math.}\ 68 (1946) 230--236.

\bibitem{Fell57}
W.~Feller,
\emph{An Introduction to Probability Theory and Its Applications},
Vol.~I, 2nd ed., Wiley, New York, 1957.

\bibitem{FlSe06}
P.~Flajolet, R.~Sedgewick,
\emph{Analytic Combinatorics},
book manuscript available at
$\bra$\url{http://algo.inria.fr/flajolet/Publications/books.html}$\ket$.

\bibitem{FHSW02}
C.~Flamm, I.L.~Hofacker, P.F.~Stadler, M.T.~Wolfinger,
Barrier trees of degenerate landscapes,
\emph{Zeitschrift f\"ur Physikalische~Chemie}
216 (2002) 155--174.

\bibitem{FrCS97}
J.~Frank, P.~Cheeseman, J.~Stutz,
When gravity fails: local search topology,
\emph{J.\ Artificial Intelligence Res.}\ 7 (1997) 249--281.

\bibitem{FLRZ01}
S.~Franz, M.~Leone, F.~Ricci-Tersenghi, R.~Zecchina,
Exact solutions for diluted spin glasses and optimization problems,
\emph{Phys.\ Rev.\ Lett.}\ 87 (2001) 127209.

\bibitem{FMRW01}
S.~Franz, M.~M\'ezard, F.~Ricci-Tersenghi, M.~Weigt, R.~Zecchina,
A ferromagnet with a glass transition,
\emph{Europhys.\ Lett.}\ 55 (2001) 465--471.

\bibitem{Frie99}
E.~Friedgut,
Sharp threholds of graph properties and the $k$-SAT problem,
\emph{J.\ Amer.\ Math.\ Soc.}\ 12 (1999) 1017--1054.

\bibitem{Gall63}
R.G.~Gallager,
\emph{Low-Density Parity-Check Codes},
MIT Press, Cambridge, Mass., 1963.

\bibitem{GnKo54}
B.V.~Gnedenko, A.N.~Kolmogorov,
\emph{Limit Distributions for Sums of Independent Random Variables},
Addison-Wesley, Cambridge, Mass., 1954.

\bibitem{GoSe05}
C.P.~Gomes, B.~Selman,
Can get satisfaction,
\emph{Nature} 435 (2005) 751--752.

\bibitem{GoCr77}
I.J.~Good, J.F.~Crook,
The enumeration of arrays and a generalization related 
to contingency tables,
\emph{Discrete Math.}\ 19 (1977) 23--45. 

\bibitem{GrMW06}
C.~Greenhill, B.D.~McKay, X.~Wang, 
Asymptotic enumeration of sparse 0-1 matrices with 
irregular row and column sums,
\emph{J.\ Combin.\ Theory Ser.\ A} 113 (2006) 291--324. 

\bibitem{HJKN06}
H.~Haanp\"a\"a, M.~J\"arvisalo, P.~Kaski, I.~Niemel\"a,
Hard satisfiable clause sets for benchmarking equivalence reasoning
techniques,
\emph{Journal on Satisfiability, 
Boolean Modeling and Computation} 2 (2006) 27--46.

\bibitem{Haje88}
B.~Hajek,
Cooling schedules for optimal annealing,
\emph{Math.\ Oper.\ Res.}\ 13 (1988) 311--329.

\bibitem{HaCK06}
J.~Hannig, E.K.P.~Chong, S.R.~Kulkarni,
Relative frequencies of generalized simulated annealing,
\emph{Math.\ Oper.\ Res.}\ 31 (2006) 199--216.

\bibitem{Haym56}
W.K.~Hayman, 
A generalisation of Stirling's formula,
\emph{J.\ Reine Angew.\ Math.}\ 196 (1956) 67--95.

\bibitem{Hirs00}
E.A.~Hirsch,
SAT local search algorithms: worst-case study,
\emph{J.\ Automat.\ Reason.}\ 24 (2000) 127--143.

\bibitem{HoHW96}
T.~Hogg, B.A.~Hubermann, C.P.~Williams (Eds.),
Frontiers in problem solving: phase transitions and complexity,
\emph{Artificial Intelligence} 81 (1996) no.~1--2.

\bibitem{HoLW06}
S.~Hoory, N.~Linial, A.~Wigderson,
Expander graphs and their applications,
\emph{Bull.\ Amer.\ Math.\ Soc.}\ 43 (2006) 439--561.

\bibitem{Hoos02}
H.~Hoos,
An adaptive noise mechanism for WalkSAT,
in: Proc.~18th National Conference on Artificial Intelligence 
(Edmonton, July 28--Aug.~1, 2002),
AAAI Press, Menlo Park, Calif., 2002, pp.~655--660.

\bibitem{HoSt05}
H.H.~Hoos, T.~St\"utzle,
\emph{Stochastic Local Search: Foundations and Applications},
Morgan Kaufmann, San Francisco, 2005.

\bibitem{Jarv06}
M.~J\"arvisalo,
Further investigations into regular XORSAT,
in: Proc.~21st National Conference on 
Artificial Intelligence (Boston, July 16--20, 2006),
AAAI Press, Menlo Park, Calif., 2006, pp.~1873--1874.

\bibitem{JiMS05}
H.~Jia, C.~Moore, B.~Selman,
From spin glasses to hard satisfiable formulas,
in: H.H.~Hoos, D.G.~Mitchell (Eds.),
Theory and Applications of Satisfiability Testing,
7th International Conference
(Vancouver, May 10-13, 2004),
Springer, Berlin, 2005, pp.~199--210.

\bibitem{KiGV83}
S.~Kirkpatrick, C.D.~Gelatt, M.P. Vecchi,
Optimization by simulated annealing,
\emph{Science} 220 (1983) 671--680.

\bibitem{KiSe94}
S.~Kirkpatrick, B.~Selman,
Critical behavior in the satisfiability of random Boolean
expressions,
\emph{Science} 264 (1994) 1297--1301.

\bibitem{LiSh02}
S.~Litsyn, V.~Shevelev,
On ensembles of low-density parity-check codes: 
asymptotic distance distributions,
\emph{IEEE Trans.\ Inform.\ Theory} 48 (2002) 887--908.

\bibitem{MaMW05}
E.~Maneva, E.~Mossel, M.J.~Wainwright,
A new look at survey propagation and its generalizations,
in: Proc.~16th ACM-SIAM Symposium on Discrete Algorithms 
(Vancouver, Jan.~23--25, 2005),
ACM Press, New York, 2005, pp.~1089--1098.

\bibitem{McSK97}
D.~McAllester, B.~Selman, H.~Kautz,
Evidence for invariants in local search,
in: Proc.~10th National Conference on Artifical Intelligence 
(Providence, R.I., July 27--31, 1997),
AAAI Press, Menlo Park, Calif., 1997, pp.~321--326.

\bibitem{MeMZ05a}
S.~Mertens, M.~M\'ezard, R.~Zecchina,
Threshold values of random $K$-SAT from the cavity method,
\emph{Random Structures Algorithms} 27 (2005) 201--226.

\bibitem{MeRR53}
N.~Metropolis, A.W.~Rosenbluth, M.N.~Rosenbluth, A.H.~Teller, E.~Teller,
Equation of state calculations by fast computing machines,
\emph{J.\ Chem.\ Phys.}\ 21 (1953) 1087--1092.

\bibitem{MeMZ05b}
M.~M\'ezard, T.~Mora, R.~Zecchina,
Clustering of solutions in the random satisfiability problem,
\emph{Phys.\ Rev.\ Lett.}\ 94 (2005) 197205.

\bibitem{MePR05}
M.~M\'ezard, M.~Palassini, O.~Rivoire,
Landscape of solutions in constraint satisfaction problems,
\emph{Phys.\ Rev.\ Lett.}\ 95 (2005) 200202.

\bibitem{MePV87}
M.~M\'ezard, G.~Parisi, M.A.~Virasoro,
\emph{Spin Glass Theory and Beyond},
World Scientific, Singapore, 1987.

\bibitem{MePZ02}
M.~M\'ezard, G.~Parisi, R.~Zecchina,
Analytic and algorithmic solution of random satisfiability problems,
\emph{Science} 297 (2002) 812--815.

\bibitem{MeRZ03}
M.~M\'ezard, F.~Ricci-Tersenghi, R.~Zecchina,
Two solutions to diluted $p$-spin models and XORSAT problems,
\emph{J.\ Stat.\ Phys.}\ 111 (2003) 505--533.

\bibitem{MiSL92}
D.~Mitchell, B.~Selman, H.~Levesque,
Hard and easy distributions of SAT problems,
in: Proc.~10th National Conference on Artifical Intelligence 
(San Jose, Calif., July 12--16, 1992),
AAAI Press, Menlo Park, Calif., 1992, pp.~459--465.

\bibitem{MZKS99}
R.~Monasson, R.~Zecchina, S.~Kirkpatrick, B.~Selman, L.~Troyansky,
Determining computational complexity 
from characteristics `phase transitions,'
\emph{Nature} 400 (1999) 133--137.

\bibitem{MoSe06a}
A.~Montanari, R.~Semerjian,
On the dynamics of the glass transition in Bethe lattices,
\emph{J.\ Stat.\ Phys.}\ 124 (2006) 103--189.

\bibitem{MoSe06b}
A.~Montanari, R.~Semerjian,
Rigorous inequalities between length and time scales in glassy systems,
\emph{J.\ Stat.\ Phys.}\ 125 (2006) 23--54.

\bibitem{MoRi04}
A.~Montanari, F.~Ricci-Tersenghi,
Cooling-schedule dependence of the dynamics of mean-field glasses,
\emph{Phys.\ Rev.\ B} 70 (2004) 134406.

\bibitem{MoMe02}
T.~Mora, M.~M\'ezard,
Random $K$-satisfiability problem: from an analytic solution to
an efficient algorithm,
\emph{Phys.\ Rev.\ E} 66 (2002) 056126.

\bibitem{MoMe06}
T.~Mora, M.~M\'ezard,
Geometrical organization of solutions to random
linear Boolean equations,
\emph{J. Stat.\ Mech.\ Theory Exp.}\ (2006) P10007.

\bibitem{MoMZ05}
T.~Mora, M.~M\'ezard, R.~Zecchina,
Pairs of SAT assignments and clustering in random Boolean formulae,
ArXiv ePrint cond-mat/0506053
$\bra$\url{http://arxiv.org/abs/cond-mat/0506053}$\ket$, 2005.

\bibitem{NeMo99}
M.E.J.~Newman, C.~Moore,
Glassy dynamics in an exactly solvable spin model,
\emph{Phys.\ Rev.\ E} 60 (1999) 5068--5072.

\bibitem{Onei69}
P.E.~O'Neil,
Asymptotics and random matrices with row-sum and column sum-restrictions,
\emph{Bull.\ Amer.\ Math.\ Soc.}\ 75 (1969) 1276--1282.

\bibitem{OrVZ05}
A.~Orlitsky, K.~Viswanathan, J.~Zhang,
Stopping set distribution of LDPC code ensembles,
\emph{IEEE Trans.\ Inform.\ Theory} 51 (2005) 929--953.

\bibitem{Papa91}
C.H.~Papadimitriou,
On selecting a satisfying truth assignment,
in: Proc.~32nd IEEE Symposium on Foundations of Computer Science
(San Juan, Puerto Rico, October 1--4, 1991),
IEEE Computer Society Press, Los Alamitos, Calif., 1991, pp.~163-169.

\bibitem{Papa94}
C.H.~Papadimitriou,
\emph{Computational Complexity},
Addison-Wesley, Reading, Mass., 1994.

\bibitem{Rath06}
V.~Rathi,
On the asymptotic weight and stopping set distribution 
of regular LDPC ensembles,
\emph{IEEE Trans.\ Inform.\ Theory} 52 (2006) 4212--4218.

\bibitem{Read59}
R.C.~Read,
The enumeration of locally restricted graphs,
\emph{J.\ London Math.\ Soc.}\ 34 (1959) 417--436 and 35 (1960) 344--351.

\bibitem{ReSt02}
C.M.~Reidys, P.F.~Stadler, 
Combinatorial landscapes,
\emph{SIAM Rev.}\ 44 (2002) 3--54.

\bibitem{RiWZ01}
F.~Ricci-Tersenghi, M.~Weigt, R.~Zecchina,
Simplest random $K$-satisfiability problem,
\emph{Phys.\ Rev.\ E} 63 (2001) 026702.

\bibitem{RiUr06}
T.~Richardson, R.~Urbanke,
\emph{Modern Coding Theory},
book manuscript available at
$\bra$\url{http://lthcwww.epfl.ch/mct/index.php}$\ket$.

\bibitem{RiKi92}
H.~Rieger, T.R.~Kirkpatrick,
Disordered $p$-spin interaction models on Husimi trees,
\emph{Phys.\ Rev.\ B} 45 (1992) 9772--9777.

\bibitem{Robb55}
H.~Robbins,
A remark on Stirling's formula,
\emph{Amer.\ Math.\ Monthly} 62 (1955) 26--29.

\bibitem{SaSF02}
P.~Salamon, P.~Sibani, R.~Frost,
\emph{Facts, Conjectures, and Improvements for Simulated Annealing},
Society for Industrial and Applied Mathematics,
Philadelphia, 2002.

\bibitem{Scha78}
T.J.~Schaefer,
The complexity of satisfiability problems,
in: Proc.~10th ACM Symposium on Theory of Computing 
(San Diego, May 1--3, 1978), ACM Press, New York, 1978, pp.~216--226.

\bibitem{Scho02}
U.~Sch\"oning,
A probabilistic algorithm for $k$-SAT based on limited
local search and restart,
\emph{Algorithmica} 32 (2002) 615--623.

\bibitem{SeAO05}
S.~Seitz, M.~Alava, P.~Orponen,
Focused local search for random 3-satisfiability,
\emph{J. Stat.\ Mech.\ Theory Exp.}\ (2005) P06006.

\bibitem{SeKC96}
B.~Selman, H.A.~Kautz, B.~Cohen,
Local search strategies for satisfiability testing,
in: D.S.~Johnson, M.A.~Trick (Eds.),
\emph{Cliques, Coloring, and Satisfiability},
American Mathematical Society, Providence, R.I., 1996, pp.~521--532.

\bibitem{SeLM92}
B.~Selman, H.~Levesque, D.~Mitchell,
A new method for solving hard satisfiability problems,
in: Proc.~10th National Conference on Artifical Intelligence 
(San Jose, Calif., July 12--16, 1992),
AAAI Press, Menlo Park, Calif., 1992, pp.~440--446.

\bibitem{SeMo03}
G.~Semerjian, R.~Monasson,
Relaxation and metastability in a local search procedure for the
random satisfiability problem,
\emph{Phys.\ Rev.\ E} 67 (2003) 066103.

\bibitem{Urqu87}
A.~Urquhart, 
Hard examples for resolution,
\emph{J.\ ACM}\ 34 (1987) 209--219. 

\bibitem{WhWa63}
E.T.~Whittaker, G.N.~Watson,
\emph{A Course of Modern Analysis}, 4th ed., 
Cambridge University Press, Cambridge, 1963.

\bibitem{Worm99}
N.C.~Wormald, 
Models of random regular graphs,
in: J.D.~Lamb and D.A.~Preece (Eds.),
\emph{Surveys in Combinatorics, 1999},
Cambridge University Press, 
Cambridge, 1999, pp.~239--298.

\end{thebibliography}
\end{document}